\documentclass[aps,prb,twocolumn,superscriptaddress,longbibliography]{revtex4-2}
\pdfoutput=1
\usepackage[english]{babel}
\usepackage{xcolor}
\newcommand{\comment}[1]{}
\usepackage{amsmath}
\usepackage{amsfonts}
\usepackage{tensor}
\usepackage{physics}
\usepackage[colorlinks=true, allcolors=blue]{hyperref}
\usepackage{siunitx}
\usepackage{booktabs}
\usepackage{bbold}
\usepackage{parskip}
\usepackage{graphicx}% Include figure files
\usepackage{dcolumn}% Align table columns on decimal point
\usepackage{bm}% bold math
\begin{document}

\title{Lamb-Dicke Dynamics of Interacting Rydberg Atoms Coupled to the Motion of an Optical Tweezer Array}
\author{Aslam Parvej}
\email{aslam.parvej@uni-hamburg.de}
\affiliation{Zentrum f\"ur Optische Quantentechnologien, Universit\"at Hamburg, 22761 Hamburg, Germany}
\affiliation{Institut für Quantenphysik, Universit\"at Hamburg, 22761 Hamburg, Germany}

\author{Ludwig Mathey}
\affiliation{Zentrum f\"ur Optische Quantentechnologien, Universit\"at Hamburg, 22761 Hamburg, Germany}
\affiliation{Institut für Quantenphysik, Universit\"at Hamburg, 22761 Hamburg, Germany}
\affiliation{The Hamburg Centre for Ultrafast Imaging, 22761 Hamburg, Germany}

\begin{abstract}
Neutral Rydberg atoms trapped in optical tweezer arrays provide a platform for quantum simulation and computation. 
In this work, we investigate the Lamb-Dicke dynamics of coupled Rydberg atoms for different trapping frequencies. 
We model the atomic motion by both internal and motional degrees of freedom, in which the motional states arise due 
to the oscillation of each atom in optical tweezer traps due to the light-atom interaction. 
In this setup, the internal states are coupled to a laser light with a Rabi frequency, while each internal state of 
each atom is also harmonically trapped with a trap frequency that depends on the internal state. 
The impact of the coherent motion of the optical tweezers on the collective dynamics of the many-body Rydberg atoms
is explored for varying Lamb-Dicke parameters and with different trap frequencies. We see the occurrence of dynamical phases e.g., 
Rabi oscillations in the decoupled limit, the limit torus phase for magic trapping, and the limit cycle phase as the 
trap frequency is further increased.
\end{abstract}
\maketitle

%--------------------------------------------------------
\section{Introduction}
%--------------------------------------------------------
\label{sec:Introduction}

In recent years, quantum simulation and quantum computation have tremendously progressed as an emerging field due to their significance in
quantum information processing, quantum cryptography, quantum metrology, and so on. There are various platforms for the 
quantum simulation and quantum computation such as trapped ions \cite{bruzewicz2019trapped, RevModPhys.93.025001}, neutral Rydberg 
atoms \cite{Yuri-2006, RevModPhys.82.2313, PRXQuantum.2.030322, ncomms15813, Ebadi-Nature}, ultra-cold molecules \cite{Carr_2009, Blackmore_2019} 
and superconducting qubits \cite{superconducting, Kaminsky2004ScalableSA}. Among them, trapped neutral Rydberg atoms in optical tweezers is 
one of the promising platforms as it is scalable with its architecture along with the lifetimes of the Rydberg atoms compared to the 
characteristics time-scale of Rydberg-Rydberg interaction and precisely controls the position and momentum of the atoms for the 
recoil-free implementation of any quantum state-preparation \cite{kley2024optimalrecoilfreestatepreparation}. 

It also provides the accessibility to explore exotic phases of the equilibrium and non-equilibrium quantum many-body systems by 
modeling with various spin models \cite{science.abi8794, s41586-023-05859-2, nature12483, s41586-020-3033-y} by encoding the ground 
state $\ket{g}$ and Rydberg state $\ket{R}$ as logical qubit space $\ket{0}$ and $\ket{1}$ along with the entangled motional degrees 
of freedom $\ket{n}$ of the optical tweezers. Recent research \cite{PhysRevLett.124.043402, PhysRevLett.125.033602, PhysRevLett.131.093002, 
PhysRevA.107.063106, PhysRevA.110.053321, Chew-Nature-Photonics, PhysRevLett.133.093405} has made great efforts to couple the Rydberg atom's 
internal and motional degrees of freedom because of their possible use in quantum many-body physics, quantum information processing, and 
quantum simulation. Over the years, these models have been utilized to investigate the Rydberg atom system in an optical 
lattice \cite{Zeiher2016-Nature-Physics, PhysRevX.8.021069, PhysRevA.108.053314}, 
in a gas phase \cite{PhysRevX.11.011011, PhysRevLett.124.063601, Takei2016, PhysRevA.76.013413},
and in an optical tweezer array \cite{science.abi8794, Chen2023, PhysRevX.14.011025, PhysRevLett.130.243001}.
%---------------------------------Figure---------------------------------------------------
\begin{figure}[b]
    % \centering
\includegraphics[width=8.5cm]{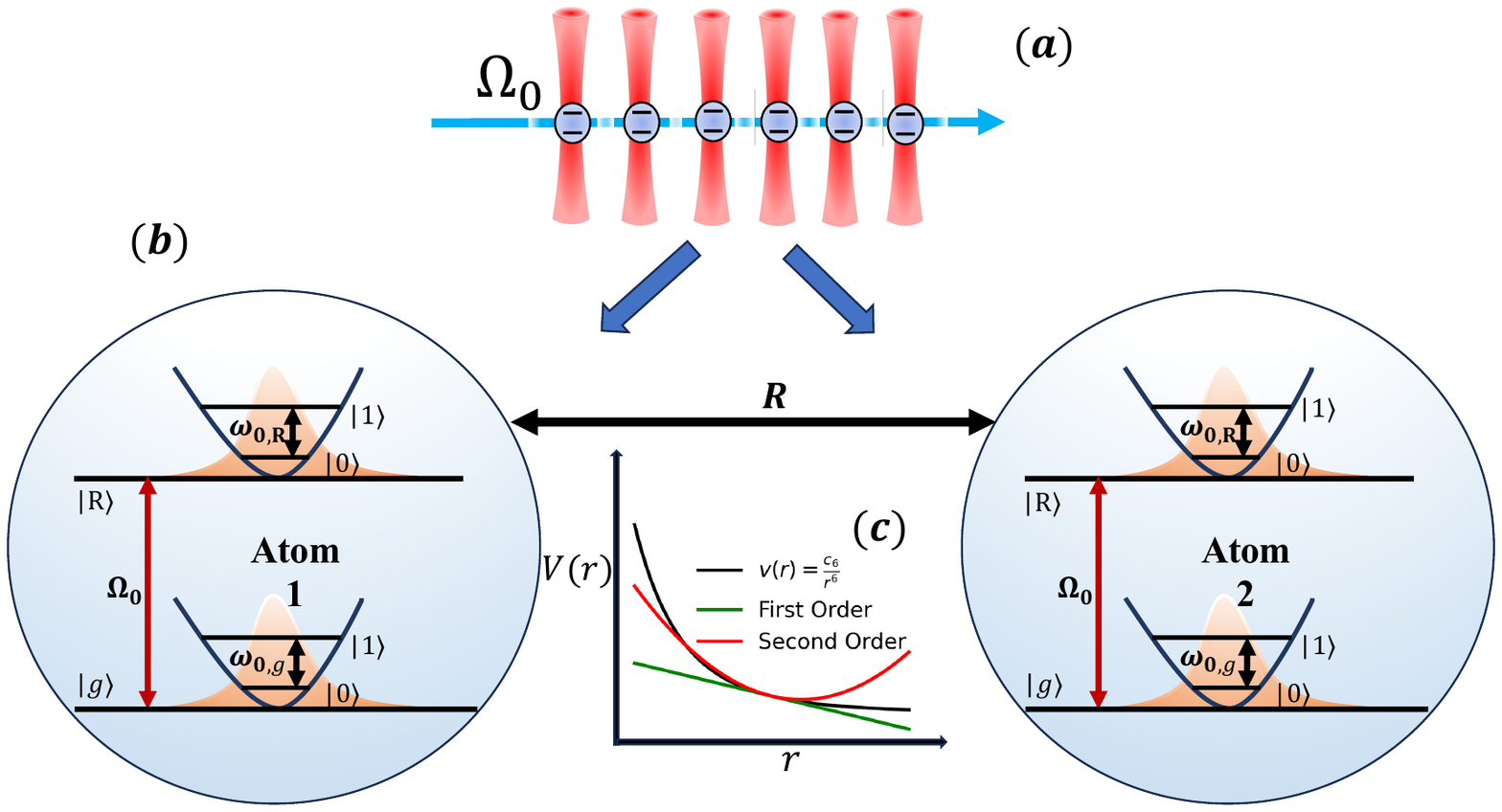}
\caption{\textbf{Lamb-Dicke coupled interacting Rydberg atoms.} Fig. (\textbf{(a)}) shows interacting Rydberg atoms 
trapped in an optical tweezer and atoms are driven by a CW laser of frequency $\Omega_0$. In Fig. (\textbf{(b)}) sketch of two atoms, 
$\textbf{(Atom1)}$ and $\textbf{(Atom2)}$ in which internal states on each atom are $\ket{g}$ and $\ket{R}$ and the motional states are 
represented by $\ket{0}$ and $\ket{1}$ by truncating to two levels of the boson states. We consider state-specific 
trapping with frequency $\omega_{0,g}$ and $\omega_{0,R}$. Two atoms are separated by a distance $R$. Atoms are interacting via van der Waals potential 
and different terms of the potential in the motional space are shown in Fig. (\textbf{(c)}).}
\label{fig:System-figure}
\end{figure}
%---------------------------------Figure---------------------------------------------------

To study the quantum many-body dynamics of our system described in Fig. (\ref{fig:System-figure}) we used the exact diagonalization (ED) method. 
The dimension of the system increases exponentially with the size of the system $N$ and uses the sparsity of the Hamiltonian, which scales as
$\sim O(N \cdot 2^{N})$ instead of $2^N \times 2^N$. For the time evolution of the entire system, the RK4 method with small time steps $dt$ is used.
%---------------------------------Figure---------------------------------------------------
\begin{figure*}[t] 
    % \centering
\includegraphics[width=\textwidth]{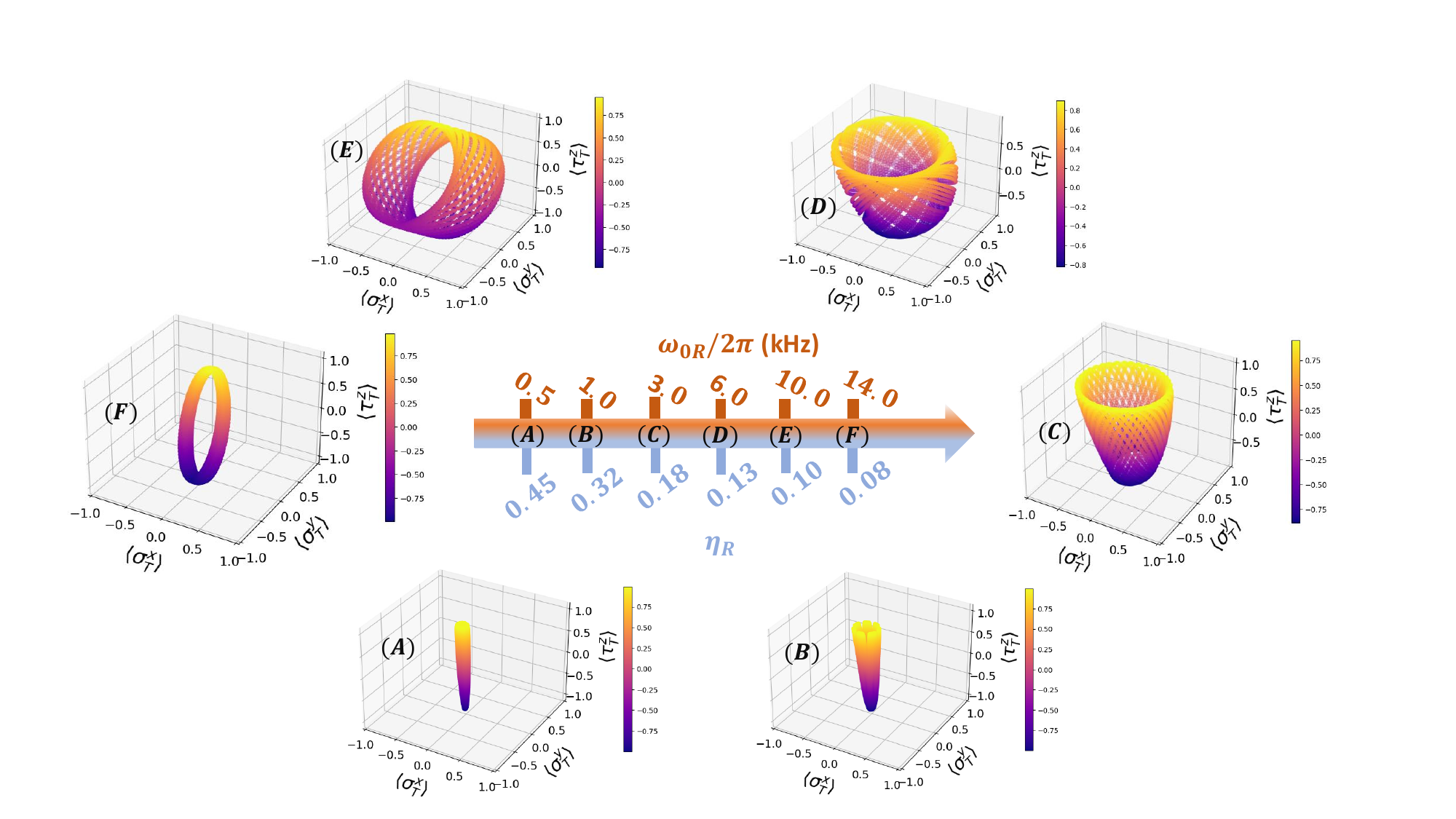}
\caption{\textbf{Dynamical phases of the interacting Rydberg atoms.} In each three-dimensional 
figure \textbf{(A)}-\textbf{(F)} the displacement $\langle \sigma_T^x\rangle$ and the 
momentum $\langle\sigma_T^y\rangle$ in the motional space of the optical tweezer, and the total density of the 
Rydberg atoms $\langle\tau_T^z\rangle$ in the internal state for different $\omega_{0,R}$ and $\eta_R$ are shown. 
The bicolor arrow at the center displays six different values of $(\omega_{0,R}, \eta_R)$ for six different phases: \textbf{(A)}$(\omega_{0,R},\eta_R)=(0.5, 0.45)$, \textbf{(B)}$(\omega_{0,R},\eta_R)=(1.0, 0.32)$, 
\textbf{(C)}$(\omega_{0,R}, \eta_R)=(3.0, 0.18)$, \textbf{(D)}$(\omega_{0,R},\eta_R)=(6.0, 0.13)$, \textbf{(E)}$(\omega_{0,R},\eta_R)=(10.0, 0.1)$, and \textbf{(F)}$(\omega_{0,R},\eta_R)=(14.0, 0.08)$ 
where $\omega_{0,R}/2\pi$ is given in unit of kHz.}
\label{fig:dynamical-phase-diagram}
\end{figure*}
%---------------------------------Figure---------------------------------------------------
    
In this paper, we present and describe the non-equilibrium dynamics in the Lamb-Dicke regime of weakly interacting Rydberg atoms 
in arrays of optical tweezers in which each atom's internal state is coupled with tweezer motion. Initially, we prepare all atoms 
in the quantum many-body ground state in both the internal and the motional degrees of freedom. Atoms are then irradiated and excited 
by the CW laser of frequency $\Omega_0$ that is ramped up adiabatically until it reaches the desired Rabi frequency. 
All atoms are trapped in a finite potential. Here, we consider state-specific trapping of each atom by trapping atoms internal Rydberg state $\ket{R}$ and 
ground state $\ket{g}$ with two different frequencies to tune the population more robustly. 

In addition, when two nearest-neighbor atoms are in Rydberg state interacting via van der Waals (vdW) potential and considering small 
oscillation about the equilibrium position of each optical tweezer by considering two harmonic motional levels and expanding this potential 
up to second order. The zeroth order represents the Rydberg-Rydberg interaction, whereas the first order represents atomic-displacement 
in the positive and negative direction by a staggered type of potential, and the second order is responsible for the hopping type of 
spin exchange mechanism. We identify three regimes based on the atom-atom interaction, the weak interaction, the blockade regime, 
very strong interacting regime. We mainly focus on the weakly interacting regime relative to the blockade radius.

This paper is organized as follows. In Sect. \ref{sec:Model and System}, we introduce the model Hamiltonian and describe the system. 
Section \ref{sec:DynamicalPhases} presents different dynamical phases of the systems in the $\eta_R$ and $\omega_{0,R}$ parameter space.
In Sect. \ref{sec:decoupled-eta-0.0}, we explain the Rabi oscillation phase for the decoupled internal and motional degrees of freedom of the 
Rydberg atoms in the optical tweezer. In Sect. \ref{sec:weakly-interacting-regime}, we demonstrate the Lamb-Dicke dynamics of 
the weakly interacting Rydberg atoms with possible experimental realizations and show the existence of limit-cycle and limit-tori phases. 
In Sect. \ref{sec:conclusion}, we discuss our results.

%--------------------------------------------------------------
\section{Interacting Rydberg atoms in optical tweezers:}
\label{sec:Model and System}
%--------------------------------------------------------------
We consider a one-dimensional (1D) chain consisting of $N$ atoms held in $N$ traps, each separated by a 
distance $R$ with each trap holding a single atom. We define the internal state of each atom by two levels, the ground 
state $\ket{g}$ and the Rydberg state $\ket{R}$. In addition, each atom has motional 
degrees of freedom due to the motion of the traps, and we consider two motional levels $\ket{0}$ and $\ket{1}$ of the harmonic oscillator levels on each atom. 
Two atoms located at sites $i$ and $j$ in the Rydberg state interact via a vdW interaction. 
Thus, the basis states of our model on each site are $\ket{g,0}$, $\ket{g,1}$, $\ket{R,0}$, and $\ket{R,1}$. The Hamiltonian of 
the full system is written as
\begin{equation}
    H(t) = H_{Rabi}(t) + H_{int} + H_{trap}
    \label{eq:full-Hamiltonian}
\end{equation}

The Rabi Hamiltonian describes the interaction of laser light with the ground state $\ket{g}$ and the Rydberg state $\ket{R}$ of 
each Rydberg atom coupled with the motional degrees of freedom in each optical tweezer and it is written in the  rotating wave approximation (RWA) as 
\begin{equation}
    \begin{split}
        % H_{Rabi} & = \frac{\hbar\Omega(t)}{2} \sum_j \left( \tau_j^+ \otimes e^{i\eta(a_j + a_j^\dagger)} + \text{h.c.} \right) \\
%         H_{Rabi} = \frac{\hbar\Omega(t)}{2} \sum_j \Big[(\text{cos}(\eta)\tau_{j}^{x}\otimes \mathbb{1} - \text{sin}(\eta) \tau_j^{y}\otimes \sigma_{j}^{x} %\Big]\\
         H_{Rabi}(t) = \frac{\hbar\Omega(t)}{2} \zeta \ \mathrm{exp\Big(-\frac{\eta_{gR}^2}{2}}\Big) \sum_j \tau_j^{+} \\ \otimes \Big[ H_{Rabi}^{m_1} + H_{Rabi}^{m_2} + H_{Rabi}^{m_3} + H_{Rabi}^{m_4} \Big]\\
        \end{split}
        \label{eq:Rabi-Hamiltonian}
\end{equation}
% \begin{equation}
%     \begin{split}
%         H_{Rabi} & = \frac{\hbar\Omega(t)}{2} \sum_j \left( \tau_j^+ \otimes e^{i\eta(a_j + a_j^\dagger)} + \text{h.c.} \right) \\
%         & = \frac{\hbar\Omega(t)}{2} \sum_j \left( \tau_j^+ \otimes e^{i\eta(\sigma_j + \sigma_j^\dagger)} + \text{h.c.} \right) \\
%          & = \frac{\hbar\Omega(t)}{2} \sum_j [(\text{cos}(\eta)\tau_{j}^{x}\otimes \mathbb{1} - \text{sin}(\eta) \tau_j^{y}\otimes \sigma_{j}^{x}]\\
%         \end{split}
%         \label{eq:Rabi-Hamiltonian}
% \end{equation}
where the Rabi Hamiltonian for the motional part $m_1, m_2, m_3,$ and $m_4$ due to the state specific traps is written as
\begin{equation}
    \begin{split}
    H_{Rabi}^{m_1} & = \frac{(1-\sigma_j^z)}{2} \\
    H_{Rabi}^{m_2} & = (1-\eta_{gR}^2)\zeta^2 \frac{(1+\sigma_j^z)}{2}  \\
    H_{Rabi}^{m_3} & = i \eta_g \zeta^2 \sigma_j^+ \\
    H_{Rabi}^{m_4} & = i \eta_R \zeta^2 \sigma_j^-
        \end{split}
        % \label{eq:Rabi-Hamiltonian}
\end{equation}

with $\eta_g$ and $\eta_R$, are Lamb-Dicke parameters for the coupling of $\ket{g}$ and $\ket{R}$, respectively, 
$\eta_{gR}$ is the effective Lamb-Dicke parameter, $\zeta$ is the ratio of the trapping frequency of the Rydberg state to the ground state, 
 and these are defined in Appendices [\ref{eq:A-Rabi-Model1} -\ref{eq:A-Rabi-Model7}].
% \begin{equation}
%     \begin{split}
%     \eta_g & = \frac{kx_{0,g}}{\sqrt{2}}  \\
%     \eta_R & = \frac{kx_{0,R}}{\sqrt{2}}  \\
%     \zeta & = \frac{\sqrt{2x_{0,g} x_{0,R}}}{\sqrt{ x_{0,g}^2 + x_{0,R}^2 }} \\
%     \eta_{gR}^2 & = \frac{k^2 x_{0,g}^2 x_{0,R}^2}{x_{0,g}^2 + x_{0,R}^2}
%         \end{split}
%         \label{eq:Rabi-Hamiltonian}
% \end{equation}

$\Omega(t)$ is the Rabi frequency which is gradually switched on following a linear 
ramp of time $t$. The motivation of this protocol is to start from the ground state of the system, and during its evolution, the system is
connected adiabatically to its ground state. The protocol for the Rabi frequency is following
\begin{equation}
{\Omega(t)} = \begin{cases}
t(\frac{r}{T}), & \text{if ${\Omega(t)} < \Omega_0$}\\
\Omega_{0},  & \text{otherwise}.
\end{cases}
 \label{eq:ramp-protocol}
\end{equation}
where $\Omega_0$ is the maximum amplitude of the Rabi frequency, $r = 2\pi \times 1$kHz is the ramp rate, 
and $T=\frac{2\pi}{\Omega_0}$ is the period of the Rabi drive. \\

We construct our system such that it allows for state-specific trapping by trapping both the 
Rydberg state $\ket{R}$ and the ground state $\ket{g}$ with frequency $\omega_{0,R}$ and $\omega_{0,g}$, respectively, and
this is written as follows
\begin{equation}
    \begin{split}
        H_{trap}^{g} & = \hbar \Big(\bar{\omega_0} - \frac{1}{2} \Delta \omega_0\Big) \sum_j \frac{(\mathbb{1}-\tau_j^z)}{2} \otimes  \sigma_j^{+} \sigma_j^{-} \\
        H_{trap}^{R} & = \hbar \Big(\bar{\omega_0} + \frac{1}{2} \Delta \omega_0\Big) \sum_j \frac{(\mathbb{1}+\tau_j^z)}{2} \otimes  \sigma_j^{+} \sigma_j^{-} \\
        H_{trap} & = H_{trap}^{g} + H_{trap}^{R} \\
        \end{split}
        \label{eq:Trap-Hamiltonian}
\end{equation}
where $\bar{\omega_0} = (\omega_{0,g}+\omega_{0,R})/2$ and $\Delta \omega = \omega_{0,R}-\omega_{0,g}$.\\

%------------------------ Figure --------------------------------------------------
\begin{figure}[b]
    \centering
    \includegraphics[width=8.5cm]{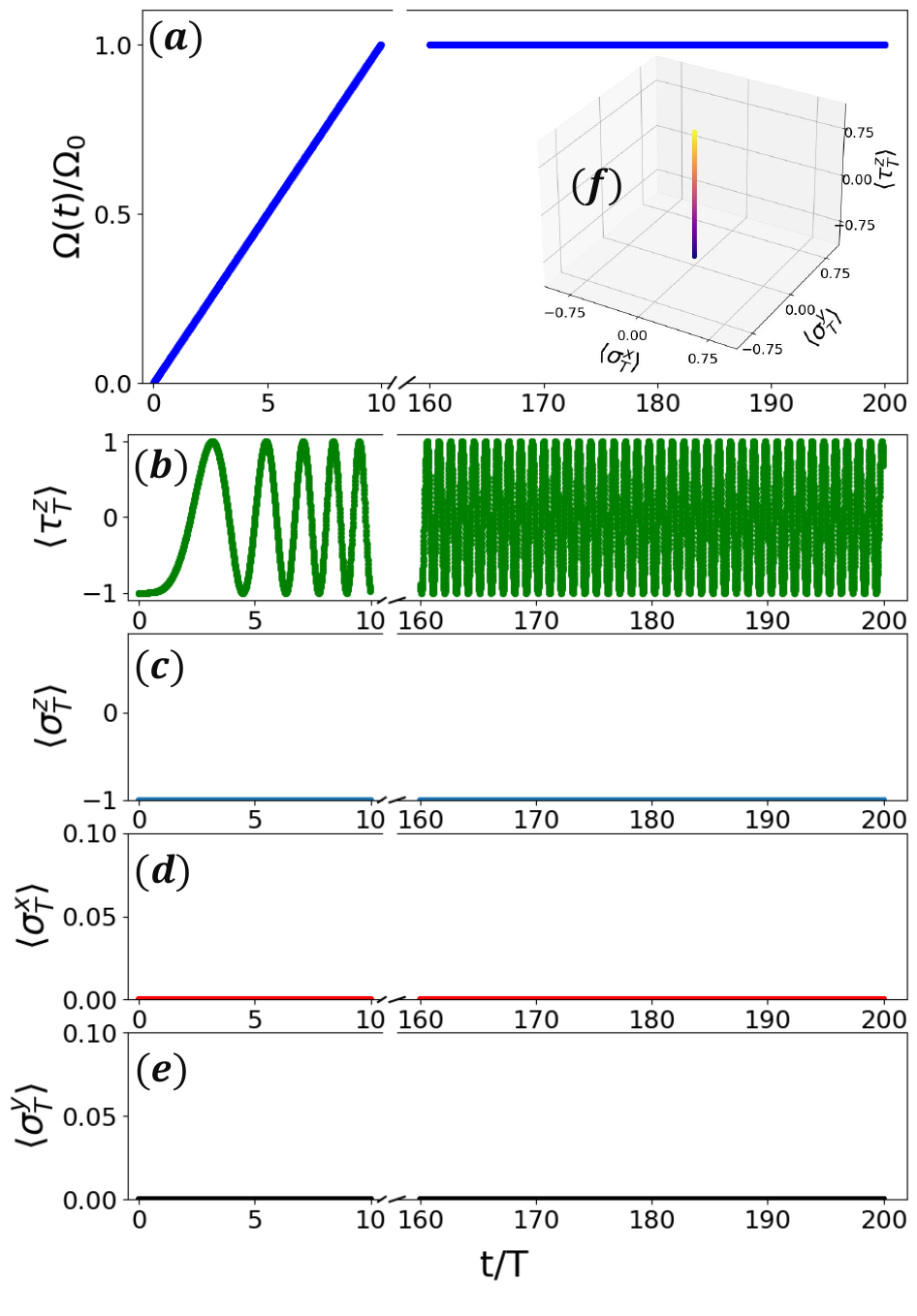}
 \caption{{\textbf{Ramp protocol, density, \& phase space trajectory}} (\textbf{a}) Protocol of Rabi frequency, 
 (\textbf{b}) population density $\langle\tau_T^z\rangle$ and (\textbf{c}) $\langle\sigma_T^z\rangle$ 
 in internal and motional spaces, respectively along with (\textbf{d}) displacement $\langle\sigma_T^x\rangle$  and (\textbf{e})
 momentum $\langle\sigma_T^y\rangle$ in motional space on the Rydberg chain in the decoupling limit ($\eta_g=\eta_R=0$) as a function of $t/T$ 
for the system size $N=20$. Inset (\textbf{f}) corresponds to the phase space of Rabi oscillations which is shown in Fig. (\ref{fig:phasespace-etag-etaR-0}) in the Appendix.}
\label{fig:total-density-eta-0.0}
\end{figure}
%------------------------ Figure --------------------------------------------------

%------------------------ Figure --------------------------------------------------
\begin{figure}[b]
    \centering
    \includegraphics[width=8.5cm]{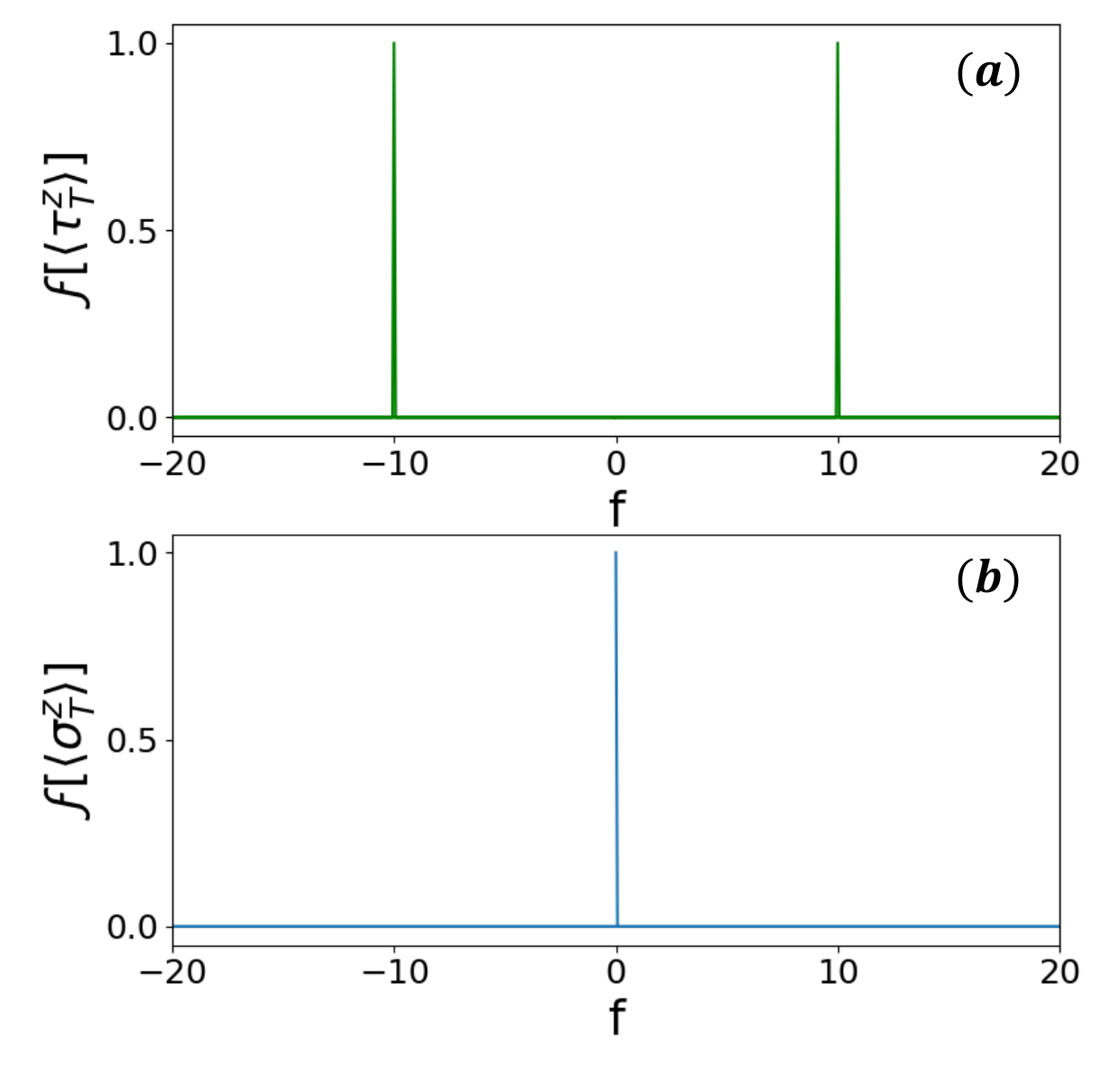}
\caption{\textbf{Discrete Fourier transform of total density.} \textbf{(a)} on the internal state and \textbf{(b)} 
on the motional state with $\eta_g=\eta_R=0$ for $N=20$ sites.}
\label{fig:fourier-eta-0}
\end{figure}
%------------------------ Figure --------------------------------------------------

Two nearest-neighbor (NN) atoms are trapped in two different traps, and when they are in the Rydberg state, they interact via the vdW potential. 
Each trap has a small oscillation and the Hamiltonian can be written in the many-body setup as
%---------------------------------------------------
\begin{equation}
    \begin{split}
        H_{int} & = \sum_{i,j \in NN} \frac{1}{4}\Bigg((\mathbb{1}+\tau_i^{z})\otimes(\mathbb{1}+\tau_j^{z}) \Bigg) \otimes P(r_{ij})
        \end{split}
        \label{eq:Interaction-Hamiltonian}
\end{equation}
%---------------------------------------------------

where $P(r_{ij})$ is due to the oscillation of each trap and expanded about the equilibrium position of each atom in optical 
tweezers up to second order and is written as 
\begin{equation}
    \begin{split}
      P(r_{ij}) & = \sum_{i,j} \Big( P_0(r_{ij}) + P_1(r_{ij}) + P_2(r_{ij}) \Big) \\
    P_0(r_{ij}) & = V_0(r_{ij}) \\
    P_1(r_{ij}) & = V_1(r_{ij}) (\sigma_{i}^{x}-\sigma_{j}^{x})  \\
    P_2(r_{ij}) &= \frac{1}{2!}  V_2(r_{ij}) (\sigma_{i}^{+}\sigma_{j}^{-}+\sigma_{i}^{-}\sigma_{j}^{+}) \\
    %   P(r_{ij}) & = \sum_{i,j} \Big( P_0(r_{ij}) + P_1(r_{ij}) + P_2(r_{ij}) \Big) \\
    % P_0(r_{ij}) & = \sum_{i,j}  V_0(r_{ij}) \\
    % P_1(r_{ij}) & = V_1(r_{ij}) (\delta r_{i}-\delta r_{j}) = V_1(r_{ij}) (\sigma_{i}^{x}-\sigma_{j}^{x})  \\
    % P_2(r_{ij}) & =\frac{1}{2!}  V_2(r_{ij}) (\delta r_{i}-\delta r_{j})^2 \\
    %             &= \frac{1}{2!}  V_2(r_{ij}) (\sigma_{i}^{+}\sigma_{j}^{-}+\sigma_{i}^{-}\sigma_{j}^{+}) 
        \end{split}
        \label{eq:Interaction-potential}
\end{equation}

The small oscillation of the atoms is written in terms of the bosonic operators as 
$\delta r_i = \sqrt{\frac{\hbar}{2m\omega_0}}(a+a^{\dagger})$, $V_0(r_{ij}) = \frac{\hbar C_6}{|r_i-r_j|^6}$, $V_1(r_{ij}) 
= \frac{-6\hbar C_6}{|r_i-r_j|^7} x_0$, $V_2(r_{ij}) = \frac{6 \times 7 \hbar C_6}{|r_i-r_j|^8} x_0^2$, 
the wave vector $k$ of the control laser, and the length of the 
harmonic oscillator $x_0$ of the harmonic oscillator. $C_6$ is the coefficient of the vdW interaction between the 
Rydberg-Rydberg states, and we choose $C_6 = 1\text{MHz}.\mu \text{m}^6$. 

As described in Eq. (\ref{eq:Rabi-Hamiltonian}), the Rydberg transition is driven by Rabi frequency $\Omega_0$ and to avoid multiple excitations we consider
the Rydberg blockade $R_b$ which is defined as $R_b = (\frac{\hbar \Omega_0}{C_6})^{-1/6}$. It allows only a single Rydberg excitation. 
Based on the blockade radius $R_b$ and Rydberg atom separation $R$, We define three distinct regimes, 
such as the regime of weak interaction ($R >>R_b$), the blockade regime ($R=R_b$) and the regime of strong interaction ($R<< R_b$). 
In this paper, we mainly focus on the weak interaction regime ($R >>R_b$).

For the experimentally realizable parameters we consider the Alkaline-Earth $\mathrm{^{171}Yb}$ atoms trapped in optical tweezers with state-specific 
trapping potential for the Rydberg state $\ket{R}=\ket{^{3}S_{1}}$ with variable frequency $\mathrm{\omega_{0,R}/2\pi = (3-10) \ kHz}$ and for the ground state
$\ket{g}=\ket{^{3}P_{0}}$ with frequency $\mathrm{\omega_{0,g}/2\pi = 10\ kHz}$ with one-photon Rabi frequency $\mathrm{\Omega_0/2\pi = 10 \ kHz}$ of the 
CW laser with wave vector $k$ varying from $\mathrm{(302-1389) \ nm}$ which corresponds to the Lamb-Dicke parameter $\eta$ varies from 0.1 to 1.0. 
The calculated value of the blockade radius $\mathrm{R_b=2.7 \mu m}$ and the typical distance between optical tweezers in $\mu m$.
% {\color{red}{For the experimentally realizable parameters we consider the Alkaline-Earth $\mathrm{^{171}Yb}$ atoms trapped in optical tweezers with state-specific 
% trapping potential for the Rydberg state $\ket{R}=\ket{^{3}S_{1}}$ with variable frequency $\mathrm{\omega_{0,R}/2\pi = (3-10) \ kHz}$ and for the ground state
% $\ket{g}=\ket{^{3}P_{0}}$ with frequency $\mathrm{\omega_{0,g}/2\pi = 10\ kHz}$ with one-photon Rabi frequency $\mathrm{\Omega_0/2\pi = 10 \ kHz}$ of the 
% CW laser with wave vector $k$ varying from $\mathrm{(302-1389) \ nm}$ which corresponds to the Lamb-Dicke parameter $\eta$ varies from 0.1 to 1.0. 
% The calculated value of the blockade radius $\mathrm{R_b=2.7 \mu m}$ and the typical distance between optical tweezers in $\mu m$}}.

%---------------------------------
\section{Dynamical Phases}
\label{sec:DynamicalPhases}
%---------------------------------
The dynamical phases of the interacting Rydberg atoms in optical tweezers presented in this work are summarized 
in Fig. (\ref{fig:dynamical-phase-diagram}) in the $\eta_R$ vs. $\omega_{0,R}$ plane shown in the middle of the figure. 
The whole phase diagram consists of a zoo of exotic phases, but in this work, we mainly focus on dynamical phases 
for different $(\eta_R, \omega_{0,R})$ parameters. These phases are Rabi oscillation, limit cycle, and limit torus and are briefly explained in the 
following section and discussed in more detail in the following sections \ref{sec:decoupled-eta-0.0}
and \ref{sec:weakly-interacting-regime}.

{\textbf{Rabi oscillation phase:}} In the Rabi oscillation phase shown 
% in Fig. \ref{fig:dynamical-phase-diagram}(A) for $\eta_g=\eta_R=0$, the population of 
in Fig. (\ref{fig:total-density-eta-0.0}) for $\eta_g=\eta_R=0$, the population of 
the Rydberg atoms oscillates with the Rabi frequency $\Omega_0$ in the internal space of the interacting Rydberg atoms 
similar to the oscillation of the two-level systems in the presence of an external Rabi drive.   

{\textbf{Limit tori phases:}} In the limit torus phase, the system shows a stable quasi-periodic oscillation 
over the long period in which the trajectory of the system in the phase space spanned by 
displacement $\langle \sigma_T^x \rangle$, momentum $\langle \sigma_T^y \rangle$
and the total density $\langle \tau_T^z \rangle$ of the Rydberg atoms in the optical 
tweezers form a torus-like structure as shown in Fig. \ref{fig:dynamical-phase-diagram}(E). 

{\textbf{Limit cycle phase:}} In this phase, the system shows a stable periodic 
oscillation and a closed trajectory in the phase space as shown in Fig. \ref{fig:dynamical-phase-diagram}(F). 
%---------------------------------
\section{Rabi Oscillation Phase in the Decoupled Limit}
% \section{Rabi Oscillation Phase in the Decoupled Limit Internal and Motional Degrees of Freedom}
\label{sec:decoupled-eta-0.0}
%---------------------------------
The quantum many-body dynamics of the whole system in the decoupled limit, i.e. $\eta_g=\eta_R=0$, where the internal state is not coupled 
with the motional state, is described in this section.

%------------------------ Figure --------------------------------------------------
\begin{figure}[t]
    \centering
    \includegraphics[width=8.5cm]{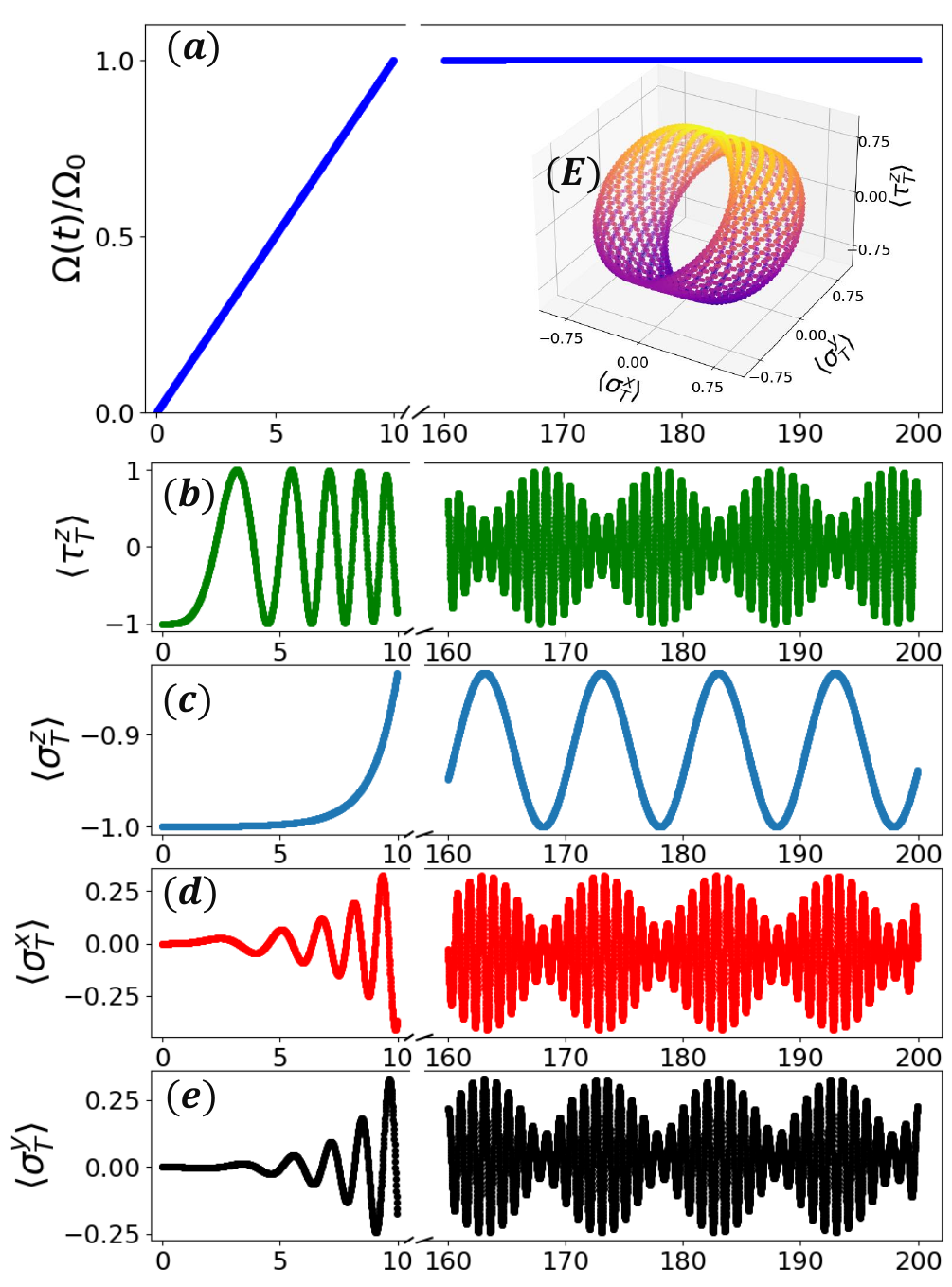}
 % \caption{{\textbf{Protocol \& population density in the internal and motional space on the Rydberg chain with ($\eta=0.1$) in the weakly interacting regime $(R/R_b)=4.0$.}} Fig. {\bf{(a)}} shows the Rabi frequency $\Omega_0(t)$ with ramp protocol starting linearly with ramp rate $r$. Fig. {\bf{(b)}} \& {\bf{(c)}} shows the total density $\langle\tau_T^z\rangle$ and $\langle\sigma_T^z\rangle$ on the internal and motional state, respectively as a function of $t/T$ with trap frequencies $\omega_{0,g}=\omega_{0,R}=2\pi \times 10 \ \text{kHz}$ for the system size $N=20$.}
  \caption{{\textbf{Ramp protocol, density, \& phase space trajectory for $\omega_{0,R}=2\pi \times 10$kHz, $\eta_R=0.1$.}} (\textbf{a}) Protocol of Rabi frequency, 
 (\textbf{b}) population density $\langle\tau_T^z\rangle$ and (\textbf{c}) $\langle\sigma_T^z\rangle$ 
 in internal and motional spaces, respectively along with (\textbf{d}) displacement $\langle\sigma_T^x\rangle$  and (\textbf{e})
 momentum $\langle\sigma_T^y\rangle$ in motional space on the Rydberg chain in the weakly interacting regime $(R/R_b)=4.0$ as a function of $t/T$ 
for the system size $N=20$.}
\label{fig:total-density-eta-0.1}
\end{figure}
%------------------------ Figure --------------------------------------------------

Each panel of Fig. \ref{fig:total-density-eta-0.0} (\textbf{a})-(\textbf{e}) shows the system dynamics over two distinct time ranges separated by
a break line. The temporal evolution of ramp protocols of the Rabi frequency from the CW lasers is shown in Fig. \ref{fig:total-density-eta-0.0} (\textbf{a}). 
At $t=0$ all atoms are in the quantum many-body ground state $\big(\ket{g}^{\otimes N/2} \otimes (\ket{0}^{\otimes N/2}\big)$ and turned on the CW laser to perform the dynamics 
linearly with ramp rate $r=2\pi \times 1$kHz until it reaches the desired Rabi frequency $\Omega_0$. The panel (\textbf{b})-(\textbf{e}) of Fig. (\ref{fig:total-density-eta-0.0}) 
shows the time evolution of the total density $\langle \tau_{T}^z \rangle$ and $\langle \sigma_{T}^z \rangle$ in the
internal and the motional spaces, respectively, the displacement $\langle \sigma_{T}^x \rangle$ and the momentum $\langle \sigma_{T}^y \rangle$ in the motional space of 
the entire system of size $N=20$ as a function of $t/T$. The transient regime is shown from time $t=0$ to $t=10T$ to reach the desired Rabi frequency $\Omega_0=2\pi \times 10 \text{kHz}$. 
The second regime is the steady-state regime from $t=160T$ to $t=200T$. We consider the time steps $dt=0.001$, the trap frequencies 
for two state dependent $\omega_{0,g}=\omega_{0,R}=2\pi \times 10 \text{kHz}$. As in $\eta_g=\eta_R=0$, the density $\langle\tau^z_{T}\rangle$ in the 
steady state shows a fast oscillation between the quantum many-body ground and the Rydberg state, whereas the total density
in the steady state shows no sideband excitation, rather it is always zero in the motional state. The total density -1 corresponds to the ground state 
density and +1 is the density of the excited Rydberg state. Similarly, there is no force acting on the tweezer motion and no momentum gain, so both these values are zero.

Fig. \ref{fig:fourier-eta-0}{\textbf{(a)-(b)}} shows the normalized discrete Fourier transform $f[\langle\tau_T^z\rangle]$ of the total 
density in the internal and motional state in Fig. \ref{fig:total-density-eta-0.0}{\textbf{(b)-(c)}}  as a 
the function of the frequency decomposition in the positive and negative spectrum for the system size $N=20$ in the decoupled limit $\eta_g=\eta_R=0$. 
We notice the presence of a dominant sharp peak at $f = 2\pi \times 10\text{kHz}$ on the internal space and a main peak at $f=0$ on the motional space. 
The occurrence of the main sharp peak of $f[\langle\tau_T^z\rangle]$ at $f = \Omega_0 = 2\pi \times 10\text{kHz}$ explains the total density has the 
fast oscillations or Rabi oscillation-like phase with this frequency in the internal states and no other sub-frequencies are present. 
%------------------------ Figure --------------------------------------------------
\begin{figure}[t]
    \centering
    \includegraphics[width=8.5cm]{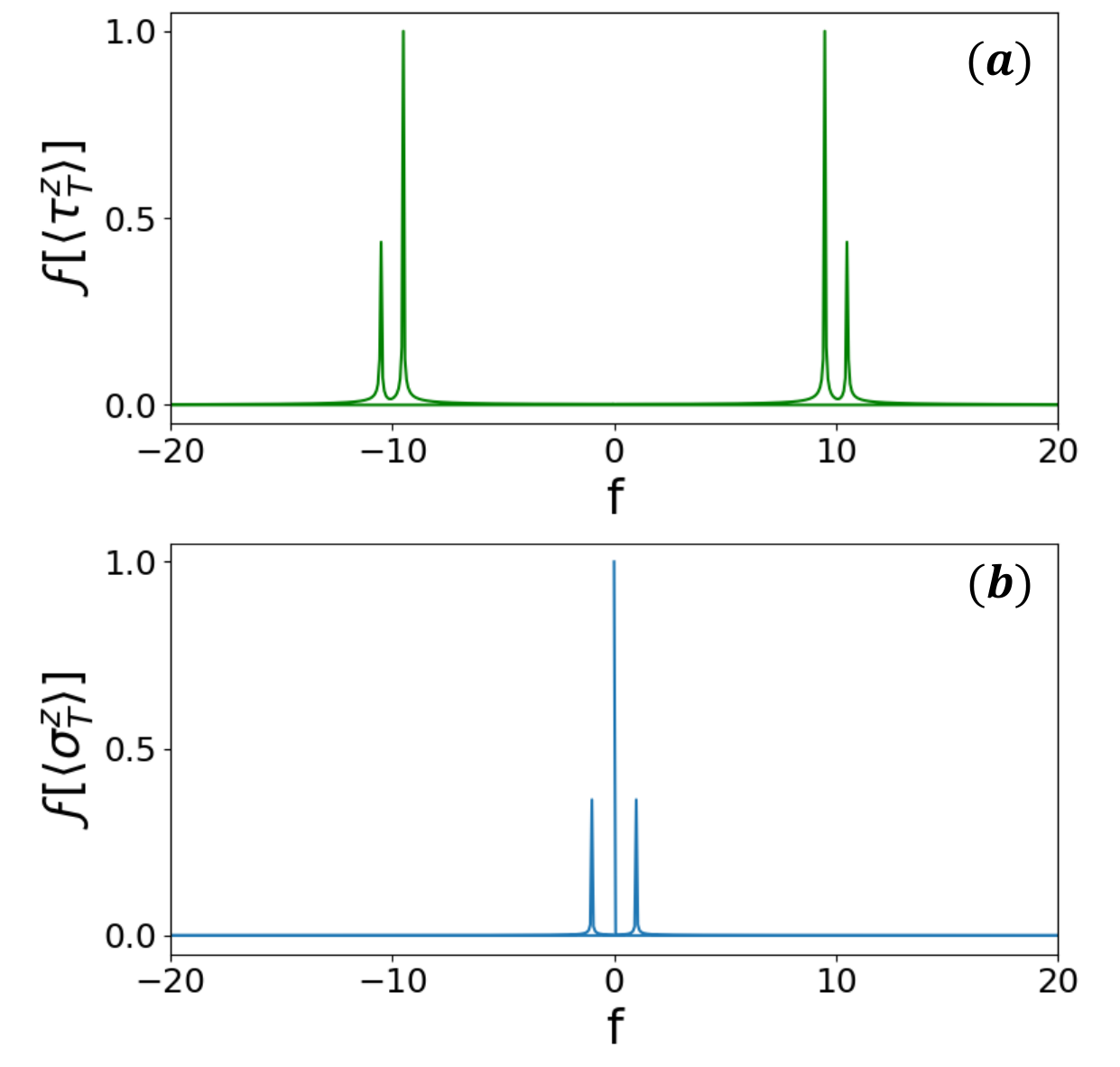}
\caption{\textbf{Discrete Fourier transform of total density.} \textbf{(a)} on the internal state and \textbf{(b)} 
on the motional state with $\eta_R=0.1$ and trap frequencies $\omega_{0,g}=\omega_{0,R}=2\pi \times 10 \ \text{kHz}$ 
for $N=20$ sites i.e., $N/2$ atoms.}
\label{fig:fourier-eta-0.1-w0R-10}
\end{figure}
%------------------------ Figure --------------------------------------------------

%-----------------------------------------
% \section{Weakly Interacting Regime}
\section{Limit Cycle and Limit Tori}
\label{sec:weakly-interacting-regime}
%-----------------------------------------
%-----------------------------------------
% \subsection{For $\eta=0.1, \ \omega_{0,R}= 2\pi\times 10 \text{kHz},\ \omega_{0,g}= 2\pi\times 10 \text{kHz}$}
% \label{subsec:w0R-10kHz}
%-----------------------------------------
We study the Lamb-Dicke dynamics of the whole system in the weakly interacting regime $R/R_b=4.0$ for 
different values of $\omega_{0,R}$ and corresponding $\eta_R$. The separation between atoms in optical 
tweezers is $R$, and the calculated value of the blockade radius is $R_b=2.15 \mu \text{m}$, which is 
within the range of experimental realization.

In Fig. (\ref{fig:total-density-eta-0.1}), we show the comparative study of (a)the Rabi protocol, 
(b)$\langle \tau_{T}^z \rangle$, (c)$\langle \sigma_{T}^z \rangle$, (d)$\langle \sigma_{T}^x \rangle$, 
and (e)$\langle \sigma_{T}^y \rangle$ of the optical tweezers in two-time intervals as a function of $t/T$. We consider
the trap frequencies for the Rydberg $\ket{R}$ and ground $\ket{g}$ states are $\omega_{0,R}=2\pi \times 10 \text{kHz}$ and 
$\omega_{0,g}=2\pi \times 10 \text{kHz}$, respectively, and corresponding Lamb-Dicke parameters are 
$\eta_R=0.1$ and $\eta_g=0.1$, respectively. The maximum Rabi-frequency $\Omega_0=2\pi \times 10 \text{kHz}$  is
shown in Fig. (\ref{fig:total-density-eta-0.1})(a). We simulate $N=20$ Rydberg atoms with $N/2$ for the internal 
and $N/2$ for the motional Fock spaces. 

Fig. (\ref{fig:fourier-eta-0.1-w0R-10}){\textbf{(a)-(e)}} demonstrates the temporal evolution of different observables 
in two time regimes, such as the transient regime for the time range $t=0$ to $t=10T$ and the long-time limit from $t=160$ to $200T$. 
We notice in the transient regime that atoms are populating the excited Rydberg states and slowly occupying the motional states. 
This happened because now the internal and motional states are coupled. 
However, in the long-time limit, the total density in the internal and motional states shows steady-state behavior. Additionally, the long-time 
dynamics on the internal state has a periodic pattern of subharmonic oscillations. 
Similarly, the motional space also shows slow pronounced oscillations with finite amplitude.

To understand the subharmonic oscillation in the internal states of the atoms, i.e., between the ground and Rydberg states, we presented the
discrete Fourier transform in Fig. (\ref{fig:fourier-eta-0.1-w0R-10}) of the total density in the internal and
motional states shown in Figs. \ref{fig:total-density-eta-0.1}(a)-(b) for $N=20$ sites. We notice a dominant and 
subdominant peak at frequency $f=10$kHz close to the Rabi frequency $\Omega_0$ in the internal space and a central 
peak in the motional space. The incommensurate ratio of these dominant and subdominant peaks in the internal space 
is the defining signature of the quasi-periodic nature or formation of a limit-torus of the coupled Rydberg tweezer 
motion system. The limit-torus is shown in Fig. \ref{fig:dynamical-phase-diagram}(E) within the phase space defined by 
displacement, momentum, and total density.

As, the value of $\omega_{0,R}$ increases the corresponding value of Lamb-Dicke parameter $\eta_R$ decreases, indicating that 
the system undergoes a phase transition from a vase-like structure Fig. \ref{fig:dynamical-phase-diagram}(A-D) 
to limit-torus Fig. \ref{fig:dynamical-phase-diagram}(E), and eventually to a limit-cycle phase Fig. \ref{fig:dynamical-phase-diagram}(F). 
The closed trajectory in the phase space is shown in Fig. \ref{fig:dynamical-phase-diagram}(F) for the discrete time $t=0$ to $t=200T$. 
The total density, displacement, and momentum for the limit-cycle phase are shown in Fig. (\ref{fig:total-density-for-all-eta-w0R}) in Appendix. 
To complement and characterize this phase we perform a discrete Fourier transform, which shows a single dominant peak whose frequency is 
other than the Rabi frequency in Fig. (\ref{fig:fourier-all-eta-woR}) in the Appendix. This is the signature of a periodic oscillation 
with a limit-cycle phase. Further increasing the value of $\eta$ to 1.0, the system remains in the limit-cycle phase 
Fig. \ref{fig:dynamical-phase-diagram}(D) shows a periodic oscillation with a frequency of oscillation that is 
different from the Rabi frequency and the trajectory is also rotated compared to the trajectory of the value $\eta=0.5$.
% %---------------------------------------------
\section{Discussion}
\label{sec:conclusion}
% %---------------------------------------------
In this paper, we have studied the Lamb-Dicke dynamics and their dynamical phases in the $\eta_R$ vs. $\omega_{0,R}$ planes of 
the coupled internal and motional degrees of freedom of Rydberg atoms in optical tweezers array in the presence of weak vdW 
Rydberg-Rydberg interaction based on the Rydberg blockade radius. We consider a model with $N/2$ interacting Rydberg 
atoms on $N=20$ sites with two internal and two motional states due to the vibration of atoms in each optical tweezer. In each atom the internal 
states $\ket{g}$ and $\ket{R}$ are trapped by two trapping frequencies $\omega_{0,g}$ and $\omega_{0,R}$, respectively. We start 
by preparing all atoms in a quantum many-body ground state $\big(\ket{g}^{\otimes N/2} \otimes \ket{0}^{\otimes N/2}\big)$ and a 
Rabi frequency is applied by a CW laser with a linear ramp to evolve the state adiabatically through the interacting Hamiltonian \ref{eq:full-Hamiltonian}. 
We use exact diagonalization (ED) to solve the Hamiltonian and the RK4 method to solve the dynamics of the Lamb-Dicke coupled system. 

The quantum many-body dynamics in the decoupled limit ($\eta_R = \eta_g=0$) show the Rabi oscillation phase. It is characterized by the 
fast oscillation of the total density in the internal state of Rydberg atoms. We focus mainly on the weakly interacting regime ($R/R_b=4.0$). 
In the small Lamb-Dicke limit, for $\eta_R = \eta_g =  0.1$ and corresponding $\omega_{0,R}=\omega_{0,g}=2\pi \times 10$kHz we observe a limit-torus phase. 
This phase is characterized by the quasiperiodic nature of the total density in the internal and motional spaces over long periods and the presence of
dominant and subdominant peaks in the discrete Fourier transform of these quantities. For $\eta_R =0.08$ and $\eta_g = 0.1$ and corresponding 
$\omega_{0,R}=2\pi \times 10$kHz and $\omega_{0,g}=2\pi \times 10$kHz we notice a phase transition to a limit-cycle phase.
In this phase, the density on the internal and motional spaces shows stable periodic oscillation shown in Fig. (\ref{fig:total-density-for-all-eta-w0R}) (F) and a dominant peak in the discrete Fourier transform of the density on the internal and motional spaces. With this study, we have expanded quantum simulation 
in Rydberg tweezer arrays towards emergent dynamical states via spin-motion coupling. We have identified limit tori and limit cycle state 
generated by the Lamb-Dicke coupling, exemplifying the rich and intriguing dynamical phenomena in this setting.

\begin{acknowledgments}
This work is funded by the Deutsche Forschungsgemeinschaft (DFG, German Research Foundation)-SFB-925 - project 170620586 and 
the Cluster of Excellence ’Advanced Imaging of Matter’ (EXC 2056) project 390715994. The project is co-financed by 
ERDF of the European Union and by ’Fonds of the Hamburg Ministry of Science, Research, Equalities and Districts (BWFGB)’.
\end{acknowledgments} 

\bibliography{reference}
%---------------------------------------------
\appendix
%---------------------------------------------
\section{Derivation of Hamiltonian in Eq.~\ref{eq:full-Hamiltonian}}
\label{app:sec-derivation}

We define the internal state of the $j^{th}$ Rydberg atom by $\tau^{\alpha}_j$, where $\alpha=x,y$ and $z$ components of the Pauli matrices and $\tau_j^{\pm}=\tau_j^{x}\pm i\tau_j^{y}$. 
The Rydberg and ground state are written as the following basis representation $\ket{R}= \begin{bmatrix} 1 \\ 0 \end{bmatrix}$ and 
$\ket{g}= \begin{bmatrix} 0 \\ 1 \end{bmatrix}$ and the operators for the internal states are represented by the following matrices
\begin{equation}
    \begin{aligned}
    \tau_j^{+} &= \begin{bmatrix}
    0 & 1\\
    0 & 0
    \end{bmatrix}, \quad
    \tau_j^{-} = \begin{bmatrix}
    0 & 0\\
    1 & 0
    \end{bmatrix}, \quad
    \tau_j^{z} = \begin{bmatrix}
        1 & 0\\
        0 & -1
    \end{bmatrix}.
    \end{aligned}
\label{eq:A-Internal-State}
\end{equation}

We also consider the motion of $j^{th}$ atom in optical tweezers by truncating the Fock space of Bosonic creation and annihilation operator to two lowest motional 
states, and it is represented by the Pauli matrices $\sigma_j^{\pm}=\sigma_j^{x}\pm i\sigma_j^{y}$ as 

\begin{equation}
    \begin{aligned}
        \sigma_j^{+} &= \begin{bmatrix}
        0 & 1\\
        0 & 0
        \end{bmatrix}, \quad
        \sigma_j^{-} = \begin{bmatrix}
        0 & 0\\
        1 & 0
        \end{bmatrix}, \quad
        \sigma_j^{z} = \begin{bmatrix}
        1 & 0\\
        0 & -1
        \end{bmatrix}.
    \end{aligned}
    \label{eq:A-Motional-State}
\end{equation}

The Hamiltonian in Eq. (\ref{eq:full-Hamiltonian}) describing the light-atom interactions in Rydberg atoms trapped in optical tweezers is 
in which each Rydberg atom has both the internal and motional degrees of freedom. The Rabi coupled Hamiltonian in 
Eq. (\ref{eq:Rabi-Hamiltonian}) is written as 

\begin{equation}
    H_{Rabi}(t) = \frac{\hbar \Omega(t)}{2} \int dx \Big( e^{ikx} \ket{g,x} \bra{R,x} +h.c. \Big)
    \label{eq:A-Rabi-Model1}
\end{equation}
where $k$ is the wave vector of the laser, and $\ket{g,x}$, and $\ket{R,x}$ are the spatial extent of the ground and Rydberg state. 
So, the spatial extent is modeled by the two lowest levels of the Harmonic oscillator, and their convention is as follows
\begin{equation}
    \psi_0(x) = \frac{1}{\sqrt{\sqrt{\pi}x_0}}e^{-x^2/2x_0^2}
    \label{eq:A-Rabi-Model2}
\end{equation}
\begin{equation}
    \psi_1(x) = \frac{1}{\sqrt{\sqrt{\pi}x_0}} \frac{\sqrt{2}x}{x_0}e^{-x^2/2x_0^2}
    \label{eq:A-Rabi-Model3}
\end{equation}
with $x_0=\sqrt{\hbar /m\omega}$. For two different oscillators, we define $x_{0,g}=\sqrt{\hbar /m\omega_g}$, and $x_{0,R}=\sqrt{\hbar /m\omega_R}$. 
For two different oscillators, the Lamb-Dicke parameters are defined as follows
\begin{equation}
    \eta_g = \frac{kx_{0,g}}{\sqrt{2}}
    \label{eq:A-Rabi-Model4}
\end{equation}

\begin{equation}
    \eta_R = \frac{kx_{0,R}}{\sqrt{2}}
    \label{eq:A-Rabi-Model5}
\end{equation}

\begin{equation}
    \zeta = \frac{x_{0,g}x_{0,R}}{\sqrt{x_{0,g}^2+x_{0,R}^2}}
    \label{eq:A-Rabi-Model6}
\end{equation}

\begin{equation}
    \eta_{gR}^2 = \frac{k^2x_{0,g}^2x_{0,R}^2}{x_{0,g}^2+x_{0,R}^2}
    \label{eq:A-Rabi-Model7}
\end{equation}

We obtain the the matrix element between $\ket{g},$ and $\ket{R}$ state with different motional levels of oscillators as follows
\begin{equation}
    \begin{split}
        \bra{R,0}H_{Rabi}\ket{g,0} & = \frac{\hbar \Omega_0}{2}R \ e^{-\eta_{gR}^2/2} \\
        \bra{R,1}H_{Rabi}\ket{g,0} & = \frac{\hbar \Omega_0}{2} i \eta_g R^3 \ e^{-\eta_{gR}^2/2} \\
        \bra{R,0}H_{Rabi}\ket{g,1} & = \frac{\hbar \Omega_0}{2} i \eta_R R^3 \ e^{-\eta_{gR}^2/2} \\
        \bra{R,1}H_{Rabi}\ket{g,1} & = \frac{\hbar \Omega_0}{2} (1-\eta_{g,R}^2)R^3 \ e^{-\eta_{gR}^2/2} \\
    \end{split}
    \label{eq:A-Rabi-Model8}
\end{equation}

% \begin{equation}
%     \begin{split}
%     H_{Rabi} & = \frac{\hbar\Omega_0}{2} \sum_j \left( \ket{g_j}\bra{R_j} e^{i\eta(a_j+a_j^{\dagger})} + \text{h.c.} \right) \\
%     & = \frac{\hbar\Omega_0}{2} \sum_i \Big(e^{i\eta(a_{i}+a_{i}^{\dagger})} \tau_i^{+} + e^{-i\eta (a_{i}+a_{i}^{\dagger})} \tau_i^{-}\Big) \\   
%     & = \frac{\hbar\Omega_0}{2} \sum_i \Big(e^{i\eta(\sigma_{i}^{+}+\sigma_{i}^{-})} \tau_i^{+} + e^{-i\eta (\sigma_{i}^{+}+\sigma_{i}^{-})} \tau_i^{-}\Big) \\
%     & = \frac{\hbar\Omega_0}{2} \sum_i \Big(e^{i\eta \sigma_{i}^{x}} \tau_i^{+} + e^{-i\eta \sigma_{i}^{x}} \tau_i^{-}\Big) \\
%     & = \frac{\hbar\Omega_0}{2} \sum_i \Big((\text{cos}(\eta) + i \text{sin}(\eta) \sigma_{i}^{x}) \tau_i^{+} \\ 
%     &+  (\text{cos}(\eta) - i \text{sin}(\eta) \sigma_{i}^{x}) \tau_i^{-}\Big) \\ \\  
%     & = \frac{\hbar\Omega_0}{2} \sum_i \Big((\text{cos}(\eta)(\tau_{i}^{+}+\tau_{i}^{-})\otimes \mathbb{1}  \\ 
%     &+ i \text{sin}(\eta) (\tau_i^{+}-\tau_i^{-})\otimes \sigma_{i}^{x}\Big)\\ \\
%         & = \frac{\hbar\Omega_0}{2} \sum_i \Big((\text{cos}(\eta)\tau_{i}^{x}\otimes \mathbb{1} - \text{sin}(\eta) \tau_i^{y}\otimes \sigma_{i}^{x}\Big)
%     \label{eq:A-Rabi-Model}
%     \end{split}
% \end{equation}

%------------------------ Figure --------------------------------------------------
\begin{figure}[b]
    \centering
    \includegraphics[width=8.5cm]{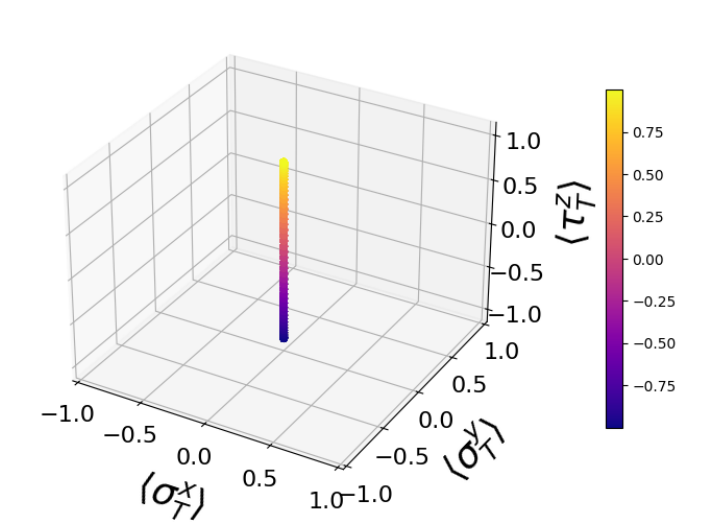}
\caption{{\bf{Phase space trajectories of interacting Rydberg atoms with $\eta_g=\eta_R=0$}}. Phase space plot shows 
the total density $\langle\tau_T^{z}\rangle$ in the internal space, momentum $\langle\sigma_T^{y}\rangle$, 
and displacement $\langle\sigma_T^{x}\rangle$ in the motional space. Trajectories along the z-direction 
show only the Rabi oscillation-like phase.}
\label{fig:phasespace-etag-etaR-0}
\end{figure}
%------------------------ Figure --------------------------------------------------
The Hamiltonian for the state-specific trap potential of Lamb-Dicke coupled Rydberg atoms in optical tweezers, with two different trap 
frequencies $\omega_{0,R}$ and $\omega_{0,g}$ corresponding to the ground and Rydberg states, respectively of the Rydberg atoms is given by the following expression
\begin{equation}
    \begin{split}
        H & = \sum_j \Big( \hbar \omega_{0,g} \ket{g_j}\bra{g_j} + \hbar \omega_{0,R} \ket{R_j}\bra{R_j}\Big) \otimes  a_j^{\dagger} a_j \\
          & =  H_{trap}^{g} + H_{trap}^{R} \\ \\
        H_{trap}^{g} & = \hbar \omega_{0,g} \sum_j \frac{(\mathbb{1}-\tau_j^z)}{2} \otimes  \sigma_j^{+} \sigma_j^{-} \\
         & = \hbar \Big(\bar{\omega_0} - \frac{1}{2} \Delta \omega_0\Big) \sum_j \frac{(\mathbb{1}-\tau_j^z)}{2} \otimes  \sigma_j^{+} \sigma_j^{-} \\ \\
        H_{trap}^{R} & = \hbar \omega_{0,R}  \sum_j \frac{(\mathbb{1}+\tau_j^z)}{2} \otimes  a_j^{\dagger} a_j \\
         & = \hbar \Big(\bar{\omega_0} + \frac{1}{2} \Delta \omega_0\Big) \sum_j \frac{(\mathbb{1}+\tau_j^z)}{2} \otimes  \sigma_j^{+} \sigma_j^{-} 
        \label{eq:A-Trap-Model}
    \end{split}
\end{equation}

where $a_i^{\dagger}$ and $a_i$ are the Bosonic creation and annihilation operator due to the motion of each Rydberg atom and we consider only two harmonic levels of the quantum oscillator with levels $n=0,1$ and can be expressed as $a_i^{\dagger}=\sigma_{i}^{+}$ and $a_i=\sigma_{i}^{-}$.

The Hamiltonian for the Rydberg-Rydberg interaction between two Rydberg atoms of the many Rydberg atoms in Eq. (\ref{eq:Interaction-Hamiltonian}) is derived from the following equation below
\begin{equation}
    % \begin{split}
        H_{int} = \sum_{i<j} V(r_i,r_j) \ket{R_i}\bra{R_i}\ket{R_j} \bra{R_j} .
    % \end{split}
        \label{eq:A-Interaction-Model}
\end{equation}

%---------------------------------Figure---------------------------------------------------
\begin{figure*}[t] 
    % \centering
    \includegraphics[width=\textwidth]{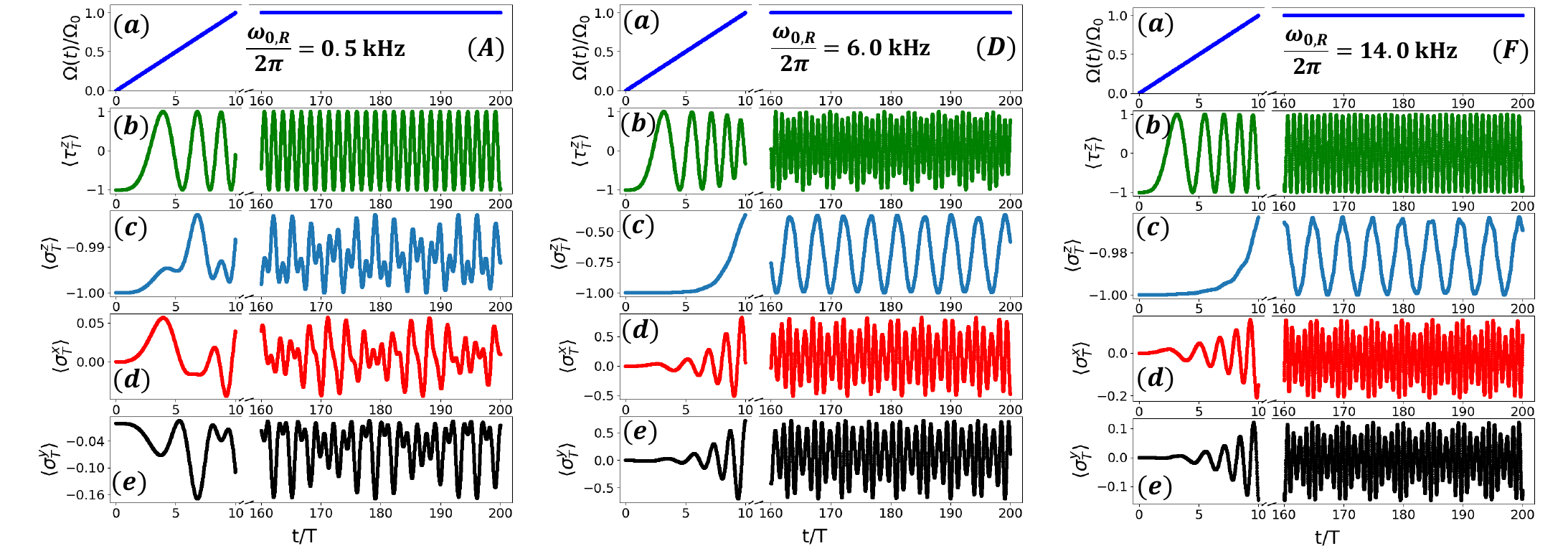}
    \caption{{\textbf{Protocol, the total density on internal and motional spaces, displacement  \&  momentum in the motional space on the Rydberg atom chain with $\omega_{0,R}=2\pi \times 10 \ \text{kHz}$.}} Fig. {\bf{(a)-(d)}} shows the Rabi frequency $\Omega_0(t)$ with ramp protocol starting linearly with ramp rate $r$. Fig. {\bf{(e)-(h)}} shows the total displacement $\langle\sigma_T^x\rangle$ \& {\bf{(i)-(l)}} the total momentum $\langle\sigma_T^y\rangle$ in the motional state as a function of $t/T$ for different $\eta =0.0, 0.1, 0.5,$ and 1.0 with trap frequency $\omega_{0,g}=2\pi \times 10 \ \text{kHz}$ for the system size $N=20$ i.e., $N/2$ Rydberg atoms in the weakly interacting regime $(R/R_b)=4.0$.}
    \label{fig:total-density-for-all-eta-w0R}
\end{figure*}
%---------------------------------Figure---------------------------------------------------
where we describe the operators with respect to the displacement $r_j$ of the $j^{th}$ atom from the center of the trap $r_j^0$. 
The separation between the two traps is $R$. Two atoms at the sites $i$ and $j$ placed in the Rydberg state interact via
the distance-dependent potential $V(r_i,r_j)=C_6/|r_i-r_j|^6$. Assume the small vibration of each atom by 
$\delta r_j=r_j-r_j^0$ about their equilibrium position and $|\delta r_j| \ll R$. The interaction potential
around the equilibrium position can be expanded and approximated up to second order as

\begin{equation}
    \begin{split}
        V({r_i,r_j}) \approx V(r_j^0,r_k^0)+\nabla V(r_j,r_k)|_{r_j^0,r_k^0} \cdot (\delta r_j,\delta r_k) \\
        \quad + \frac{1}{2} (\delta r_j,\delta r_k)^T \cdot \nabla^2 V(r_j,r_k)|_{r_j^0,r_k^0} \cdot (\delta r_j,\delta r_k) \\
       \quad \approx V_0(r_{ij})\mathbb{1} + V_1(r_{ij}) x_0 (r_i-r_j) + \frac{1}{2!} V_2(r_{ij}) x_0^2 (r_i-r_j)^{2}
    \end{split}
        \label{eq:A-Potential-Model}
\end{equation}

 The time-evolved quantum many-body wavefunction is written as

\begin{equation}
    \ket{\psi(t)} = \sum_{j=1}^{2^N} c_{ij}(t) \ \ket{\phi_i} \otimes \ket{\chi_i}
\end{equation}

The density on the internal space can be derived as follows 
\begin{equation}
    \langle \tau_{T}^z (t) \rangle_{\phi} = \frac{1}{N} \int_{t=0}^{t_f} \ \sum_{j=1}^{2^N} \bra{\psi_j(t)} \tau_{j}^z \ket{\psi_j(t)} \ dt
\end{equation}
Similarly, the density on the motional space is constructed as follows
\begin{equation}
    \langle \sigma_{T}^z (t) \rangle_{\chi} = \frac{1}{N} \int_{t=0}^{t_f} \ \sum_{j=1}^{2^N} \bra{\psi_j(t)} \sigma_{j}^z \ket{\psi_j(t)} \ dt
\end{equation}
where $\ket{\phi_j}$ $\ket{\chi_j}$ are wavefunctions on the internal and motional subspace of the system.

The displacement of atom tweezer is expressed by Pauli matrix as we have restricted each motional space to two levels as following
\begin{equation}
        \langle \sigma_{T}^x (t) \rangle_{\chi} = \frac{1}{N} \int_{t=0}^{t_f} \ \sum_{j=1}^{2^N} \bra{\psi_j(t)} \sigma_{j}^x \ket{\psi_j(t)} \ dt
\end{equation}
and the momentum of the atom tweezer due to the atom-light interaction is written as
\begin{equation}
        \langle \sigma_{T}^y (t) \rangle_{\chi} = \frac{1}{N} \int_{t=0}^{t_f} \ \sum_{j=1}^{2^N} \bra{\psi_j(t)} \sigma_{j}^y \ket{\psi_j(t)} \ dt
\end{equation}

\begin{figure*}[t] 
    % \centering
    \includegraphics[width=\textwidth]{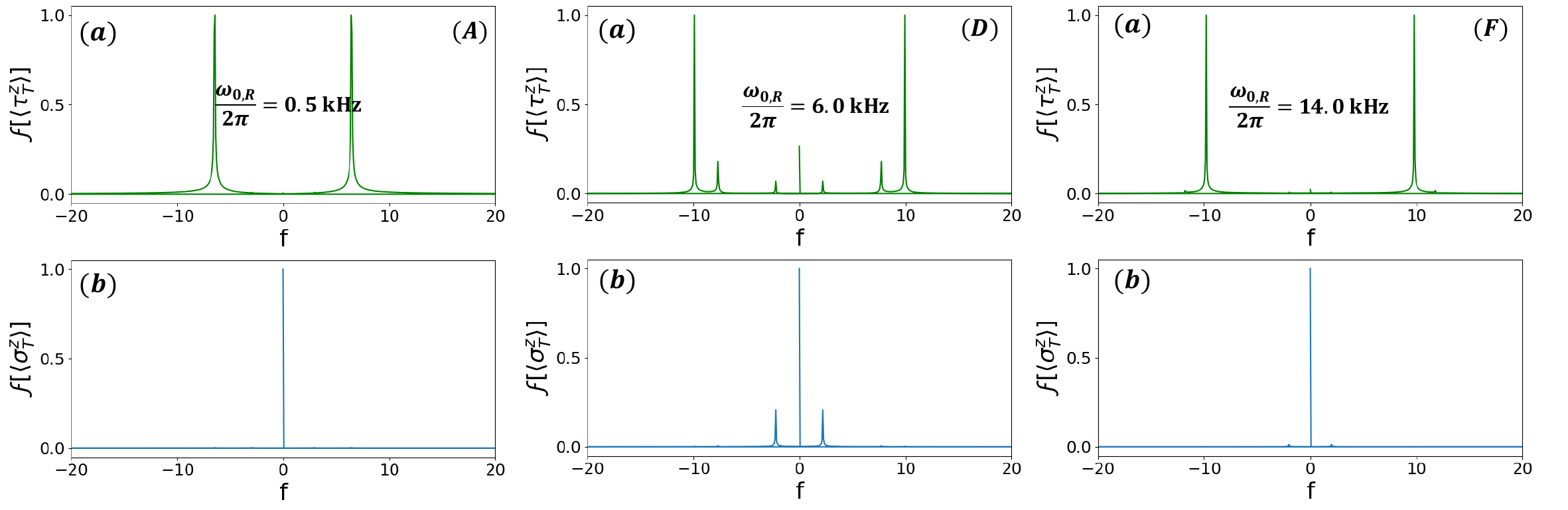}
\caption{\textbf{Discrete Fourier transform of total density.} Fig. {\bf{(a)-(c)}} on the internal state and {\bf{(d)-(f)}} 
on the motional state with $\eta=0.1$ and trap frequencies $\omega_{0,g}=\omega_{0,R}=2\pi \times 10 \ \text{kHz}$ 
for $N=20$ sites i.e., $N/2$ atoms.}
\label{fig:fourier-all-eta-woR}
\end{figure*}
%---------------------------------Figure---------------------------------------------------

% The trap potential of each atom is approximated by two levels and can be written as

% \begin{equation}
%     H_{trap} = \hbar \omega_0 \sum_i a_i^{\dagger}a_i = \hbar \omega_0 \sum_i \sigma_i^{+}\sigma_i^{-}
% \end{equation}
\end{document}